# On Deliberate Misreferencing as a Tool of Science Policy

F. Hoyle & N.C. Wickramasinghe

Dept. of Applied Mathematics and Astronomy,
University College, Cardiff, U.K.

In the second half of February the impending Giotto encounter of 13 March concentrated our minds on what the encounter might reveal. As an outcome, we issued a fairly widely circulated preprint with the title "Some Predictions on the Nature of Comet Halley." (1 March, 1986, Cardiff Series 121) whose contents were reported in the issue of the *Times* for 12 March. This publication in the *Times* was fortunate for us, because it appeared indisputably ahead of the encounter, whereas a contemporaneous submission to the Royal Astronomical Society has suffered long delays to acceptance on advice to the Society from two persons of unknown identities.

Now, more than two months later, a letter entitled "Predicting that Comet Halley is dark" by J.M. Greenberg has appeared in the issue of *Nature* for May 1986. The communication begins: "As I write this letter in early March only the scantiest information has come out of the Giotto mission." No precise date at which Professor Greenberg thus took up his pen is given.

Our predictions turned out to have the rather out-of-focus quality that normally goes with attempting to see in advance what is going to happen. We were lucky to be substantially right in two of our shots (including the prediction that the comet is dark) and probably, although not certainly we think, wrong on the third. Greenberg's predictions, on the other hand, have a startling accuracy on which we must congratulate him, since we could not have achieved anything so precise unless we had known Giotto events in advance of their occurrence.

Despite our admiration for Professor Greenberg's sibylline achievements, we think most people require predictions to be announced publicly in advance of events. Otherwise there would be no point at all in making predictions. No ordinary bookmaker, for example, pays out money to would-be clients who seek to back a winner two months after a race has been run. In science, however, bookmakers gladly pay out money even years after a race has been run, especially it seems if the client happens to be Professor Greenberg.

In the special supplement to the issue of *Nature* for 15 May there are two examples, both concerning the likely organic nature of the bulk of cometary dust. The communication "Composition of Comet Halley dust particles from Giotto observations" (J. Kissel *et al*, p 336) attributes the suggestion that cometary dust might be organic to an article by Greenberg dated 1979, while the communication "Composition of Comet Halley dust particles from Vega observations" also references Greenberg, through an article dated 1982. Our own first paper on organic material in comets appeared in 1975 (*Astrophys. and Space Science,* volume 33, L 19-28). By 1979 we had written about a dozen further papers on the subject as well as three full length books. Our views by 1979 had indeed become so widely known as to justify the word 'notorious'. Evidently then, the misreferencing in the issue of *Nature* for 15 May could not possibly have been inadvertent.

Unfortunately these examples do not stand alone - they can be matched by more instances than can be numbered on the fingers of both hands. Otherwise we would surely have followed the Christian advice to turn the other cheek. But when one has to do it more less continuously, turning the other cheek can become a painful business, about which Christian advice appears to be confused.

Most of the situations involve the journal *Nature*. In at least one instance known to us there was an inhouse deletion of a reference to our work, raising the possibility that much of the misreferencing might be due to the journal itself rather than to authors. Why *Nature* should behave in this way is a problem we have puzzled over for more than half a decade, especially as up to 1977 our relations with the journal were good.

Misreferencing would not be in question if our views from 1975 onwards have been wholly wrong. It is just because we have sometimes been right that misreferencing has arisen. The present Editor, John Maddox, an old friend, is a sufficiently experienced science journalist to know that a degree of success, as for instance our longstanding prediction of the organic nature of comets, justifies one being granted a hearing on other still controversial issues. It does not justify a policy of suppression and obliteration.

Our opposition to the Darwinian theory lies we believe at the root of the matter. At a microlevel, where only single base-pair changes of DNA are involved, the Darwinian theory is correct. But at a macrolevel involving multiple base-pair changes the theory is wrong. The latter statement can be proved both from direct facts and from extensive mathematical analysis. Evolution does not, and cannot from the mathematics, work according to Darwin at taxonomic levels from the orders of a class upwards. Evolution that is Darwinian proceeds mostly at the lowest level of all, the level of varieties, just as it was conceived to do already in the 1850's. Beyond what was then apparent, the theory does not do much. But even if all this was not clearly demonstrable it would be usual in science, in physical science at any rate, for us to be granted the right to our own opinion, instead of being exposed to an orchestrated campaign of calumny and misreferencing.

It is our view furthermore that the issue is not one over which the editorial staff of *Nature* has real control, for we suspect it to be a management condition that persons judged to be a threat to Darwinism are to be blotted out from the scientific world. Darwin is known to have been highly influential in the decision to found *Nature*. From its inception, *Nature* has acted as an enthusiastic vehicle of propaganda for Darwin's theory, behaving towards more opponents more and more as time has gone on in the fashion of a Praetorian Guard. The effect by now is that hardly a whisper of dissent from Darwinism is heard among biologists, despite the serious and increasing difficulties of that theory. It happened more or less as a fluke that we entered this century-old mental prisonhouse from astronomy, daring because of our background in mathematics and physics to speak out. To be as charitable as it is possible to be, one might possibly say that after so many years of operation the system is now running itself, without there being anybody at all in control of it.

There is no law that misreferencing shall not be used as an instrument of science policy. Nor is there a law outside the courts which requires even an approximation to the truth to be spoken or written. *Nature* is free to follow whatever course it wishes so long as scientists continue to favour the journal with their work and continue to purchase it. We are content that this should be so, for we know of no better arrangement for the publication of science than the present one. Yet it should be remembered that cultures of high quality are more fragile than they appear at first sight. While they were still operative there seemed no limit to the productions of the Elizabethan dramatists, the Florentine painters and the Viennese musicians. But all are gone now and no amount of effort, desire or money will bring them back. There is no more likely cause of a similar decline and virtual disappearance of productive science it seems to us than a calculated disrespect for the truth.

While *Nature* as a commercial enterprise may consider itself exempt from the worst of our criticisms, the same cannot be said for organisations funded out of the public purse. The policy of misreferencing used by *Nature* was adopted by the Edinburgh Royal Observatory in "Laboratory and Observational Infrared Spectra of Interstellar Dust", circulated in 1984. Although the first publications on a fair fraction of the topics discussed were contributed by us, we were unable to discover in this Edinburgh production a single citation of any of our pioneering papers. Since at least some of our papers have become widely known and, are quoted in other places, nothing but a policy of deliberate misreferencing can explain this absence from extensive reference lists running in total to several hundred entries. That such things were done at public expense is not acceptable. Nor is it acceptable in our opinion that an Observatory with disrespect for truth and accuracy should be just the one to which the Science and Engineering Research Council proposes to move the Royal Greenwich Observatory, although with a cynicism not unknown in such circles one might say that the ancient traditions of the one Observatory might at least do something to moderate the activities of the other.